\documentclass[11pt,preprint]{aastex}
\usepackage{epsfig,color}
\usepackage{mathrsfs}
\usepackage{ulem}

\def\mdot{\dot{m}}
\def\mdotss{\dot{m}_{_{\rm SS}}}
\def\Mdot{\dot{M}}
\def\bhm{M_{\bullet}}
\def\civ{C {\sc iv}}

\def\ergs{\rm erg~s^{-1}}
\def\etaSS{\eta_{_{\rm SS}}}
\def\etaSlim{\eta_{_{\rm Slim}}}
\def\fblr{f_{_{\rm BLR}}}
\def\feii{Fe {\sc ii}}

\def\hb{H$\beta$}

\def\ledd{L_{\rm Edd}}

\def\lbol{L_{\rm Bol}}
\def\Lbol{L_{\rm Bol}}
\def\kms{km~s$^{-1}$}
\def\mbh{M_{\bullet}}
\def\mgii{Mg {\sc ii}}

\def\oiii{[O {\sc iii}]}
\def\pp{\prime\prime}
\def\rblr{R_{_{\rm BLR}}}

\def\sunm{M_{\odot}}
\def\msun{$M_{\odot}$}
\def\vfwhm{V_{_{\rm FWHM}}}

\begin{document}

\title{Supermassive black holes with high accretion rates in active galactic nuclei: \\
I. First results from a new reverberation mapping campaign}

\author{
Pu Du\altaffilmark{1},
Chen Hu\altaffilmark{1},
Kai-Xing Lu\altaffilmark{2},
Fang Wang\altaffilmark{3},
Jie Qiu\altaffilmark{1},
Yan-Rong Li\altaffilmark{1},
Jin-Ming Bai\altaffilmark{3},\\
Shai Kaspi\altaffilmark{4},
Hagai Netzer\altaffilmark{4}, and
Jian-Min Wang\altaffilmark{1,5,*}\\
(SEAMBH collaboration)}

\altaffiltext{1}
{Key Laboratory for Particle Astrophysics, Institute of High Energy Physics,
 Chinese Academy of Sciences, 19B Yuquan Road, Beijing 100049, China.}

\altaffiltext{2}
{Astronomy Department, Beijing Normal University, Beijing 100875, China.}

\altaffiltext{3}
{Yunnan Observatory, Chinese Academy of Sciences, Kunming 650011, Yunnan, China.}

\altaffiltext{4}
{Wise Observatory, School of Physics and Astronomy, Tel-Aviv University, Tel-Aviv 69978, Israel}

\altaffiltext{5}
{National Astronomical Observatories of China, Chinese Academy of Sciences,
  20A Datun Road, Beijing 100020, China.}

%\altaffiltext{*}
%{Corresponding authors}

\begin{abstract}
We report first results from a large project to measure black hole (BH) mass in high
accretion rate active galactic nuclei (AGNs). Such objects may be different from other
AGNs in being powered by slim accretion disks and showing saturated accretion
luminosities, but both are not yet fully understood.
The results are part of a large reverberation mapping (RM) campaign
using the 2.4-m Shangri-La telescope at the Yunnan Observatory in China. The goals
are to investigate the gas distribution near the BH and the properties of the central
accretion disks, to measure BH mass and Eddington ratios, and to test the feasibility
of using such objects as a new type of cosmological candles. The paper
presents results for three objects, Mrk\,335, Mrk\,142 and IRAS F12397+3333
with H$\beta$ time lags relative to the 5100\AA\ continuum of $10.6^{+1.7}_{-2.9}$,
$6.4^{+0.8}_{-2.2}$ and $11.4^{+2.9}_{-1.9}$ days, respectively. The corresponding BH
masses are $(8.3_{-3.2}^{+2.6})\times 10^6\sunm$, $(3.4_{-1.2}^{+0.5})\times 10^6\sunm$
and $(7.5_{-4.1}^{+4.3})\times 10^6\sunm$, and the lower limits on the Eddington
ratios 0.6, 2.3, and 4.6 for the minimal radiative efficiency of 0.038. Mrk\,142 and
IRAS\,F12397+333 (extinction corrected) clearly deviate from the currently known relation
between H$\beta$ lag and continuum luminosity. The three Eddington ratios are beyond
the values expected in thin accretion disks and two of them are the largest measured
so far among objects with RM-based BH masses. We briefly discuss implications for slim
disks, BH growth and cosmology.
\end{abstract}


\section{Introduction}

Active galactic nuclei (AGNs) are powered by accretion onto supermassive black
holes (SMBHs) located at the centers of their host galaxies. Most of the information
about the physical conditions in the vicinity of the BH is obtained through spectroscopy
of such sources. In addition, the temporal behaviors of the continuum and emission
lines in such objects provide useful information about the distribution and emissivity
of the line emitting gas and, through reverberation mapping (RM), a way to measure
the mass of the central BH (%Bahcall \& Kozlovsky 1969;
Bahcall et al. 1972;
Blandford \& McKee 1982; Netzer 1990; Peterson 1993; Netzer \& Peterson 1997; Peterson
2013 and references therein). RM experiments, based on the response of
the broad emission lines to the continuum variations, have provided
reliable estimates of the lag between the continuum and the broad emission line light
curves in about 50 AGNs (e.g., Kaspi et al. 2000; Peterson et al 2004; Kaspi et al.
2005; Bentz et al. 2009a,b; Denney et al. 2010; Bentz et al. 2013). This information, 
combined with a measure of the gas velocity in the broad line region (BLR) was used to 
obtain an estimate of the BH mass. This involves expressions of the type,
\begin{equation}
\bhm=\fblr\frac{\rblr v^2}{G},
\end{equation}
where $\mbh$ is the BH mass, $\fblr$ is a factor that includes information about the
geometry and kinematics of the BLR gas, $\rblr$ is the responsivity weighted radius
of the BLR for the emission line in question, and $v$  is a measure of the velocity in
the line emitting gas, e.g., the full width at half maximum (FWHM) ($\vfwhm$) or the ``line
dispersion'' ($\sigma_{\rm line}$, see Peterson et al. 2004 and references therein).
For a virialized BLR with close to a spherical geometry, $\fblr \sim 1$ for velocity
characterized by $\vfwhm$ (e.g., Netzer and Marziani 2010)  and $\fblr \sim 5.5 $ for
velocity characterized by $\sigma_{\rm line}$
(Park et al 2012; Woo et al. 2013 and references therein).
The only practical way to calibrate the value of $\fblr$ is in AGNs with measured stellar
velocity dispersion in the bulge of the galaxy, $\sigma_*$, through comparison with BH mass
estimates based on the well-established $\bhm-\sigma_*$ relationship in non-active galaxies
(Ferrarese et al. 2001; Onken et al.
2004; Park et al. 2012; Woo et al. 2013). The data presented in Woo et al.
(2013), suggests that $\fblr$ does not depend on the Eddington ratio.
This includes the 7 narrow line Seyfert 1 galaxies (NLS1s) in that sample: Mrk 110,
Mrk 202, Mrk 590, Mrk 766, NGC 4051, NGC 4748 and NGC 7469.

All RM campaigns show tight $\rblr-L$ relations, where $L$ is a measure of $\lambda L_{\lambda}$ 
in the optical-UV part of the spectral energy distribution (SED), for both  H$\beta$ and the
\civ$\lambda 1549$ emission lines (Kaspi et al. 2007; Denney et al. 2013; Bentz et al. 2013). Such 
relationships provide an indirect way to estimate the BLR size, and hence $\mbh$, from ``single epoch"
spectra assuming virialized BLRs. Such methods are routinely used in studies of BH mass and
accretion rate distributions in large AGN samples such as the Sloan Digital Sky Survey (SDSS,
see e.g., McLure \& Dunlop 2004; Wang et al. 2006, 2009; Netzer et al. 2007; Netzer \& Trakhtenbrot
2007; Hu et al. 2008; Vestergaard \& Osmer 2009; Shen 2009; Trakhtenbrot \& Netzer 2012; Kelly
\& Shen 2013). The
more reliable measurements of $\mbh$ based on RM experiments, are available for only about 50
AGNs with directly measured $\rblr$. Most of these sources are ``slow accretors'', i.e., their
luminosities are well below the Eddington limit.

In this paper we assume that all BHs are powered by accretion via a central,
optically thick accretion flow that takes the shape of an accretion disk (AD).
We use the same general notation for all such flows, where
\begin{equation}
\mdot=\frac{L_{\rm Bol}}{L_{\rm Edd}}=\frac{\eta\dot{M}c^2}{L_{\rm Edd}}
=\eta\dot{\mathscr{M}},
\end{equation}
is the ``Eddington ratio''. Here, $L_{\rm Bol}$ and $L_{\rm Edd}$ are the bolometric
and Eddington luminosity, respectively, $\Mdot$ is the mass accretion rate through
the disk, $\eta$ is the efficiency of converting accreted mass to radiation and
$\dot{\mathscr{M}}=\Mdot c^2/L_{\rm Edd}$ is the ``Eddington rate''. The Eddington
luminosity we use assumes solar composition gas and is given by
$\ledd=1.5\times 10^{45}\left(\mbh/10^7\sunm\right)\ergs$.

In principle, the Eddington ratio $(\mdot$) is a measurable property obtained from the
measured $\mbh$ and the known SED.
The Eddington rate is a more theoretical concept
that is related to the physics and geometry of the accretion flow. The radiation of
geometrically thin ADs, like those based on the black body disk model
of Shakura and Sunyaev (1973; hereafter SS disk) depend only on the location of the
last stable orbit (Bardeen et al. 1972). Thus, $\etaSS\approx (0.057, 0.32, 0.038)$
for spin parameters of $a=(0, 0.998,-1)$, respectively. However, thicker ``slim disks''
are subject to radial advection which can significantly reduce the efficiency. Here,
$\etaSlim\propto \dot{\mathscr{M}}^{-1}$ and is only weakly dependent on the BH spin.
The canonic models of slim disks (e.g., Wang \& Zhou 1999a; Wang \& Netzer 2003) suggest
a logarithmic relation between $\mdot$ and $\dot{\mathscr{M}}$ in such systems. For
example, in such cases $\mdot\approx (2-3)$ for $1\lesssim\dot{\mathscr{M}}\lesssim 200$
(Abramowicz et al. 1988; Wang et al. 1999b
Mineshige et al. 2000). In principle, $\mdot$ can be determined by observations but
the determination of $\dot{\mathscr{M}}$ is more problematic since it depends on the specific
slim disk model. This has important implications to the process of cosmic growth
of BHs, the duty cycle of activity, etc.

A recent paper by Wang et al. (2013) suggests that super-Eddington accreting massive
black holes (SEAMBHs), those objects with $\dot{\mathscr{M}} \gg 1$ and 2-10 keV photon index
($\Gamma_{2-10}$) larger than 2, can be used  as a new type of cosmological
candles.\footnote{We avoid the more commonly used term "standard candle" because the 
method is based on the idea that a different BH mass results in a different luminosity and 
the sources in question have different masses.} The method is based on the fact that the bolometric 
luminosity of slim accretion disks around massive BHs tends to saturate at super-Eddington rates.
Under such conditions, $L_{\rm Bol} \propto \mbh$ with only a weak logarithmic
dependence on $\dot{\mathscr{M}}$. This idea, if confirmed by more accurate measurements, can
provide a useful tool for measuring cosmological distances at high redshifts, where
supernovae (SNe) are hard to detect or are rare (Hook 2012) because of their slow
evolution (Kobayashi \& Nomota 2009). SEAMBHs are very luminous, and can be easily
detected in large numbers up to very high redshifts (Netzer and Trakhtenbrot 2013).
In addition, their properties may not depend in any way on galaxy evolution. However,
there are a very small number of such sources with accurately measured BH mass in the 
local universe and a systematic study is necessary to test these  ideas.

We have started a large observational project, in China, aimed at increasing considerably
the number of high$-\mdot$ systems with good RM  measurements that will allow reliable
estimates of $\mbh$. The main scientific goals are:
1) Understand the physical properties of slim accretion disks.
2) Probe the physics and dynamics of the BLR gas in very fast accretors.
3) Improve the understanding of co-evolution
of galaxies and their central BHs in extreme cases where mass accretion, and
hence BH growth, are very fast.
4) Use the improved mass and accretion rate measurements to calibrate
SEAMBHs as cosmological candles.

In this  paper, the first in a series, we outline the overall program, explain
the observational aspects and introduce our first results for three BHs that are
very fast accretors. In \S2 we describe the target selection, the observational
setup and the data reduction procedure. In \S3 we present the first results of
the project and  draw preliminary conclusions. Throughout this work we assume a
$\Lambda$CDM cosmology with $H_0=67\,{\mathrm{ km~s^{-1}Mpc^{-1}}}$,
$\Omega_m=0.32$ and $\Omega_{\Lambda}=0.68$ in light of Planck observations
(Ade et al. 2013).

\section{Observations and reduction}
\subsection{Target selections}
The most important criterion for the selection of targets is based on their normalized
accretion rate as estimated by the (less accurate) single-epoch method. In particular, we
search for objects with $\mdot\ge 1$. As explained, the suggestion is that in such
cases, the radiative efficiency $\eta$ is no longer a simple function of the BH spin and
advection becomes an important process that affects the source luminosity. We decided to
avoid radio loud AGNs although, in a few cases, a radio loud source was discovered
after the project had begun. Such cases will be discussed separately.

Selecting AGNs by their high accretion rates depends on their $\Lbol$ which in itself
depends on the SED. Unfortunately, in standard ADs, much
of the radiation is emitted  beyond the Lyman limit at 912\AA\, (Laor and Netzer 1989;
Elvis et al. 1994; Marconi et al. 2004; Richards et al. 2006; Davis and Laor 2011;
Elvis et al. 2012). According to the standard SS model, the optical bolometric correction
factor, $\kappa_{\rm B}$, i.e., the factor converting $\lambda L_{\lambda}$ measured at
optical wavelengths to $\Lbol$, depends on the accretion rate and BH mass such that
$\kappa_{\rm B}\propto (\dot{m}/\mbh)^{1/3}$. For ADs around $10^8-10^9\sunm$ black holes,
this agrees with global empirical estimates that suggest $\kappa_{\rm B}\sim 5-10$
(e.g., Marconi et al. 2004). However, much larger and much smaller values of $\kappa_{\rm B}$
are predicted by thin AD models for smaller and larger BHs, respectively (see specific
examples in Netzer and Trakhtenbrot 2013). Since most of the $\sim 50$ measured BH masses
in the RM campaign have BH masses that fall below $10^8$\msun, we suspect that $\mdot$ in
at least several of these sources has been underestimated in the past. Thus, at least
some of these sources can be SEAMBHs.

Alternatively, we can select targets based on the slope of their 2-10 keV X-ray continuum.
As argued in Wang et al. (2013), some SEAMBH models predict that such objects would show
steeper X-ray slopes compared with low-$\mdot$ sources. While a possible $\mdot-\Gamma_{2-10}$
relation has been suggested by Wang et al. (2013), this depends on the poorly understood
disk-corona structure. The underlying physics is based on the link between the accretion
rate in the cold gas phase, the increased emission from the UV and optical bands, and the
enhancement of Comptonization cooling, leading to a suppressed hot corona emission. A
correlation between $\mdot$ and $\Gamma_{2-10}$ has indeed been reported in both low and
high luminosity AGNs (see e.g., Wang et al. 2004; Risaliti et al. 2009; Zhou \& Zhang 2010;
Shemmer et al. 2006; Brightman et al. 2013). A sample of 60 SEAMBH candidates selected
from the literature was presented by Wang et al. (2013). The $\mdot-\Gamma_{2-10}$
correlation in this sample is strong but with a large scatter. Much of the scatter can
be the result of the uncertain BH masses and bolometric luminosities of these sources.

Our target selection  follows several criteria that remove the dependence of the disk
structure on the uncertain $\lbol$:
\begin{enumerate}
\item
The sources are NLS1s suspected to be SEAMBHs. Their $\Lbol$ and $\mbh$ are based on the
single epoch mass determination method and satisfy $\mdot>1$. Their Eddington rates are
based on the SS thin disk model and hence\footnote{Davis and Laor (2011), following Collin
(2002) and others, derived an expression for estimating $\Mdot$ which is basically identical
to Eqn.~\ref{eq:mdot}. They showed that this simple expression agrees with more elaborated
thin AD models and used it to derive spin-independent values of $\Mdot$ based on the
4861\AA\ luminosity. They further obtained $\eta_{ss}$ by integrating the observed
and unobserved SEDs. They found that about a  third of their PG QSOs accrete at
super-Eddington rates.}:
\begin{equation}
%\dot{m}_{_{\rm SS}} \simeq 2.0\left(\frac{L_{44}}{\cos i}\right)^{3/2} M_7^{-2}\eta_{_{0.1}} \, ,
\dot{m}_{_{\rm SS}} \simeq 20.1\left(\frac{L_{44}}{\cos i}\right)^{3/2} M_7^{-2}\etaSS \, ,
\label{eq:mdot}
\end{equation}
where $L_{44}=\lambda L_{\lambda}/10^{44}\ergs$ at $\lambda=5100$\AA, $M_7=\mbh/10^7\sunm$
%, $\eta_{_{0.1}}=\etaSS/0.1$ 
and $i$ is the inclination angle. We require that $\mdotss>1$ for
$\cos i =0.75$ (an inclination  typical of type-I AGNs that cannot be observed
from a much larger inclination angle due to obscuration by the central torus). This expression
contains the unknown value of $\eta$ due to the
unknown BH spin ($a$). However, a lower limit on $\mdotss$ can be obtained by assuming
the lowest possible efficiency of thin ADs, $\etaSS=0.038$, corresponding to a spin parameter
of $a=-1$. We should point out, again, that $\mdot_{_{\rm SS}}$ given by Eqn.~\ref{eq:mdot}
is a lower limit for the Eddington ratios $\mdot$ if BH accretion proceeds via a slim AD.

\item
The targeted S/N is high enough to enable 2D velocity-resolved RM  in
the H$\beta$ and \mgii\ $\lambda 2798$ lines.

\item The 2-10 keV photon index is very steep, $\Gamma_{2-10}>2$.
\item Targets can be followed from the ground for a period of at least 80 to 180
days for low and high redshift AGNs, respectively.

\item Objects with proven continuum variations.
We note that there is some evidence that the variability amplitudes of AGNs are anti-correlated
with Eddington ratios (e.g., Ai et al. 2013; Zuo et al. 2012). In particular, some NLS1s show
small optical continuum variations (Giannuzzo et al 1996, 1998), but some are very small (Shemmer
\& Netzer 2000; Klimek et al. 2004; Yip et al. 2009). However,
significant variations are found in many SDSS NLS1s located in the Strip 82 region (Ai et al.
2013). We chose sources from the Catalina
database\footnote{http://nesssi.cacr.caltech.edu/DataRelease/} and excluded candidates showing
variations of less than 10\% in the Catalina light curves.
\end{enumerate}

\subsection{Photometric and spectrophotometric observations}
\subsubsection{The Shangri-La telescope and spectrograph}
All the spectroscopy and imaging observations reported here were obtained with
the Shangri-La telescope (SLT: IAU site code O44) at the Lijiang Station of the
Yunnan Observatory of the Chinese Academy of Sciences. The SLT started its
operation in 2008. This is a 2.4 m alt-azimuth mounted Ritchey-Chr$\mathrm{\acute{e}}$tien
telescope with a field de-rotator enabling to position two objects along the same long slit.
The RMS pointing error is about 2 arcsec rms, and the tracking
accuracy with autoguiding is better than $0.5^{\prime\prime}/$hour.
The longitude of the station is $100^{\circ}~01^{\prime}~51^{\prime\prime}$ E, the latitude
$26^{\circ}~42^{\prime}~32^{\prime\prime}$ N, and the altitude 3193 m. The site has two
observing seasons: a rainy season that lasts from June to mid-September, with very little
clear time, and a dry season, from mid-September until May, when most nights are clear.
The annually averaged seeing is $\sim 1^{\pp}.5$ in terms of the FWHM of stars (measured
with YFOSC), ranging from $0^{\pp}.7$ to $2^{\pp}.0$.

The YFOSC (Yunnan Faint Object Spectrograph and Camera), built in 2010 by the astronomical
instrumentation team at the Niels Bohr Institute, is similar to the EFOSC (ESO Faint Object
Spectrograph and Camera), but with an additional focal reducer. It started its operation
in 2011. YFOSC is a versatile instrument
for low resolution spectroscopy and imaging, working at the Cassegrain focus. The CCD chip
is an e2v CCD42-90 Back Illuminated Deep Depletion $2048\times4608$ pixel Scientific
CCD Sensor whose pixel size is 13.5 mm, pixel scale $0.283^{\prime\prime}$ pixel$^{-1}$,
covering a $10^{\prime}\times 10^{\prime}$ field of view.
Switching from photometry to spectroscopy is done automatically and takes less than one
second. The spectrograph is equipped with a large number of grisms with different dispersions
and can be used in a long slit mode as well as with several fixed apertures.

\subsubsection{Spectrophotometry}

Our RM campaign  started in October 2012. We focused on objects whose $L_{5100}$ indicates
\hb\ lag of up to 30 days as judged by the expression given in Bentz et al. (2009a). The
typical sampling frequency is very high and for most sources we managed to obtain high
quality spectra almost every night.  Observations of a particular object were terminated
when a clear line-to-continuum lag was detected and measured.
The mean number of nights per source was about $90$ and the mean monitoring
period about 8 times  the H$\beta$ delay.

We took advantage of the long slit capability of the YFOSC and observed a nearby standard
star along the slit for all objects. As explained in detail in the original papers adopting
this method (Maoz et al. 1990; Kaspi et al. 2000), this ensures high accuracy relative flux
measurements even during times of relatively poor observing conditions. Selection of
comparison stars is based on the images of the SDSS and the typical separation between
targets and comparison stars is $\sim 1^{\prime}-3^{\prime}$. Given the seeing conditions,
the entrance slit was fixed at projected width of $2^{\pp}.5$.

All the spectra were obtained using YFOSC with Grism 14 which provides a resolution of 92
\AA\,mm$^{-1}$ (1.8\AA\,pixel$^{-1}$) and covers the wavelength range of $3800 - 7200$\AA.
For each object/comparison star, we obtained three consecutive exposures in order to
remove cosmic-rays and estimate the systematic flux calibration errors. Standard neon
and helium lamps are used for wavelength calibration. The spectroscopic data are reduced
using standard IRAF v2.16 routines before absolute flux calibration. All spectra are
extracted in a uniform, rather large extraction aperture of 30 pixels ($\sim$ 8.5 arcsec)
to avoid light losses.

The flux calibration is done in two steps: (1)
Absolute flux spectra of the comparison stars are generated using the observations of
spectrophotometric standards in several nights of good weather conditions. This results
in fiducial fluxed spectra for all comparison stars. (2) For each object/comparison star
pair, a sensitivity function is obtained by comparing the spectrum of the star to the
fiducial spectrum. This produces a sensitivity function that is applied to calibrate the
observed AGN spectrum. This step resembles the IRAF {\it sensfunc} and {\it calibration},
and it also resembles the method of AGN/star ratio that was used by Maoz et al. (1990)
and Kaspi et al. (2000).

\subsubsection{Photometric observations}
We also made photometric observations through a Johnson $V$-filter of all targets and
comparison stars. This allows us to test the quality of the comparison star calibration
and the continuum light curves. The observations are done just before the spectroscopic
observations with typical exposure times of  4--5 minutes for a $m_V=15-16$ target.

The images are reduced using standard IRAF procedures. We perform
differential photometry of the targets relative to several other stars in the same field.
The number of comparison stars is typically $3-4$. The radius of the aperture photometry
is $10^{\pp}$ and the typical photometric accuracy 1--2\%.

\subsection{Line and continuum light curves}
\subsubsection{light curve measurements}

We use two continuum bands, 4760-4790\AA\, and 5085-5115\AA,
in the rest frame, to set the continuum underneath the H$\beta$ emission line.
We then integrate the continuum-subtracted H$\beta$ flux between 4810 and 4910\AA.
The red limit of the band is chosen to exclude the
\feii\, line at around 4924\AA, which is strong in some of the objects.
Two kinds of error bars are calculated for the continuum flux. First, for the
consecutive exposures in the same night, after flux calibration by the
comparison star described above, the mean flux of each exposure over the
entire wavelength range is calculated. Then the mean fluxes of exposures are
divided by that of the combined spectrum in that night to yield flux ratios.
The largest deviation of the ratios from unity is used as an error bar. Second,
the Poisson error of the measured continuum flux is calculated from the combined
spectrum in the corresponding continuum band. Typically the two error bars are
comparable (both $\sim 1\%$), but there are exceptions where one of the two
dominates. The two error bars are summed in quadrature as the final uncertainty of
the continuum flux. For the error bar of the emission line flux, only Poisson
noise is calculated. The first kind of error, difference between consecutive
exposures, includes the change in the host galaxy
contamination. The emission line fluxes should be unaffected by such changes. 
Thus, the difference between consecutive exposures is not inherited.

The above error bars on the continuum flux do not account for systematic uncertainties
that can be caused by poor weather conditions, bright moon, telescope tracking
inaccuracies, slit positioning, etc. These are manifested as flux differences
between adjacent nights that are significantly larger than the mean continuum
variations over this period (the variability time scale for all sources is much
longer than one day). To estimate these uncertainties we first smooth the continuum
light curve with a median filter of 5--6 points. We then subtract this median-smoothed
light curve from the original light curve and obtained the standard deviation from the
residuals. This serves as an estimate of the systematic uncertainty. The systematic
uncertainties are usually larger than those associated with the measurements in one
night and are therefore the largest contributors to the total errors that go into the
CCF analysis and the calculated time-lag uncertainties (there are several exceptions
where errors on individual points are larger than the systematic uncertainties). This
method is very similar to the one described by Peterson et al. (1998) and Bentz et al.
(2009).

We constructed line and continuum light curves for all of the sources. Some of the
photometric observations are seriously influenced by moon light when a target is close to
the moon. While the spectroscopic light curves follow the photometric ones, the agreement
between the two is influenced by host galaxy contamination. This is the result of the constant
slit width in all observations that may result in different host galaxy contributions on different
nights. The leading factors are seeing variations, inaccurate centering of the target and
inaccurate tracking. The host galaxy contribution can change slightly from one night to the
next, which can affect the AGN light curve.

\subsubsection{Variability characteristics }
To characterize the continuum variability of the sources we use the variability
characteristic $F_{var}$ defined by Rodr\'iguez-Pascual et al. (1997),
\begin{equation}
F_{\rm var}=\frac{\left(\sigma^2-\Delta^2\right)^{1/2}}{\langle F\rangle} \, ,
\end{equation}
where $\langle F\rangle=N^{-1}\sum_i^NF_i$ is the averaged flux, $N$ is
the number of observations and
\begin{equation}
\sigma^2=\frac{1}{N-1}\sum_i^N\left(F_i-\langle F\rangle\right)^2;~~~~
\Delta^2=\frac{1}{N}\sum_i^N\Delta_i^2.
\end{equation}
Here $\Delta_i$ represents the uncertainty on the flux $F_i$. Below we
apply Eqn (4) to the $V$-band, $F_{\rm H\beta}$ and $F_{5100}$ light
curves.

\subsubsection{Cross correlation analysis}
We employed two standard methods to analyze the correlation between the line and
continuum light curves: the interpolated cross-correlation function (ICCF, Gaskell \&
Sparke 1986; Gaskell \& Peterson 1987) and the  Z-transformed discrete correlation
function (ZDCF, Alexander 1997). The latter is an improvement on the discrete correlation
function (DCF) of Edelson \& Krolik (1988). The results of the two are in
excellent agreement and we only quote the time lags obtained with the ICCF.

The delay of the emission lines relative to the continuum variations is determined
either by measuring the location of the peak  of the
cross-correlation function (CCF) ($r_{\mathrm{max}}$) or the centroid of the points
around the peak above a certain threshold, e.g., 80\%  ($r\ge0.8r_{\mathrm{max}}$).
We note the two by $\tau_{\mathrm{peak}}$ and $\tau_{\mathrm{cent}}$. The
uncertainties on $\tau_{\mathrm{peak}}$ and $\tau_{\mathrm{cent}}$ were
calculated with the ``flux randomization(FR)/random subset sampling (RSS)''
method. This procedure is described in Peterson et al. (1998; 2004) and will not
be repeated here.

\subsection{Host galaxy contamination}
In order to obtain the Eddington ratios (Eqn. 3), we have to subtract the contaminations
of host galaxies. Following the scheme described in Bentz et al. (2009a), we estimated
the host galaxy contribution to the observed AGN flux at 5100\AA\, using archival
{\it Hubble Space Telescope} ({\it HST})
images that are available for all our  targets. In a case of more than one {\it HST}
observation, we chose those images with the longest exposure times in the optical band.

The procedure used here is quite standard and its main features are summarized here for
clarity. We retrieve from the archive the calibrated data that were reduced by the latest
version of the pipeline, and the best reference files. All of the
exposures for a single object are combined with {\it AstroDrizzle} v1.0.2 task
in the {\it DrizzlePac} package to clean cosmic rays and make one
distortion-corrected image of the host galaxy\footnote{{\it AstroDrizzle} is the new
software for aligning and combining {\it HST} images. It was officially
released in June 2012 to replace the widely used {\it Multidrizzle} task.}.
Only the pixels flagged as good in the data quality frames are included in the
combined images. Some exposures include saturated pixels. These were removed prior to
the final combination of all data. Thus, all pixels marked with zeros in weighting
{\it Astrodrizzle} output, (i.e., no good pixel from any exposure) are masked out in the
following fitting processes. Examples of the final combined distortion-free images are
shown in the left panels of Figure 1 with spectroscopic monitoring apertures
overlaid. The diagram confirms that galaxy contribution to the observed fluxes
at 5100\AA\ is non-negligible in all three cases.

We use GALFIT v3.0 (Peng et al. 2002;
2010) to model the surface brightness distribution of the three host galaxies.
The GALFIT algorithm  fits two dimensional analytic functions to galaxies and
point sources. Each of the objects in this study is fitted
with a point-spread function (PSF) to model the AGN, one to several S\'{e}rsic
profiles to model the host galaxy, and a constant to model the sky background.
PSFs were generated for each AGN by creating a distorted PSF model at its detector
position in each of the exposed frames using TinyTim package (Krist 1993) and
combining these models by {\it Astrodrizzle} using the same configuration that
was used to combine the AGN images. S\'{e}rsic profiles are employed to model the
various host galaxy components such as bulges, disks and bars. Thus the surface brightness
is expressed as
\begin{equation}
\Sigma(R) = \Sigma_e \exp \left\{ -\kappa_0 \left[ \left( \frac{R}{R_e}
\right)^{1/n} - 1 \right] \right\} \, ,
\end{equation}
where $\Sigma_e$ is the surface brightness at the effective radius $R_e$, $n$ is the
S\'ersic power-law index and $\kappa_0$ is coupled to $n$ to make sure that half of the flux
is within $R_e$ (see details in Peng et al. 2002; 2010) ($n=1$ represents a galactic
disk and $n=4$ a bulge). We set no constraint on $n$ when fitting bulge, bar or
elliptical host, but fix $n=1$ to model the disk component.

At least one S\'{e}rsic
component was required to model the galaxy in addition to the PSF and sky
components, and more were added if needed. For some cases where the PSF modeled
by TinyTim does not match the nuclear surface brightness distribution  well (see
Krist 2003; Kim et al. 2008), an additional S\'{e}rsic component with small
effective radius was used to modify the mismatch. When the field of view
(FOV) of the {\it HST} instrument [e.g., High Resolution Channel (HRC) of the
Advanced Camera for Surveys (ACS)] is small, the area in the edge of the image
used to constrain the fitted sky value was limited. We carefully adjusted the
initial parameters to make sure the residuals and $\chi^2$ were minimized.
Other bright objects in the FOV of our targets were masked out.

Having completed the host galaxy modeling, we subtracted the PSF and sky
components from the images of each object and extracted from the
PSF-sky-free images the total counts due to the host galaxy inside 
the slit of the spectrograph. These counts were transformed to flux density units.
 Because of the difference between
the pivot wavelength of the {\it HST} filter and the rest-frame 5100\AA,
we used a bulge template spectrum (Kinney et al. 1996) to
determine the required color correction. The template is redshifted and
reddened by Galactic extinction based on the dust extinction map of Schlafly \&
Finkbeiner (2011). The {\it synphot} package is then employed to convolve
the {\it HST} passband with the template to simulate the photometry.

Normalizing the counts to the observed total counts in the aperture provides
the host galaxy flux at  5100\AA. We adopted a nominal 10\% uncertainty on
the host galaxy contributions due to
the modeling procedures (see also analysis in Bentz et al. 2013).

\section{Results and discussion}
\subsection{General results}

In this first paper of the series we report the observations of three objects:
Mrk\,142, Mrk\,335 and IRAS\,F12397+3333. Information on the three targets and comparison
stars is given in Table 1 as well as the variability parameters ($F_{\rm var}$)
for the continuum and the H$\beta$ emission line. Table 2 gives the data for
the continuum and H$\beta$ light curves. Mrk 142 was previously mapped by the LAMP
collaboration (Bentz et al. 2009b). Mrk\,335 has been monitored in 2 earlier
campaigns (Kassebaum et al. 1997; Grier et al. 2012) and IRAS F12397+3333
is reported here for the first time.  Table 3 and
Fig.~1 present data and images related to the host galaxy treatment in the
three sources (see captions for details). The {\it HST} images of Mrk 142 and
Mrk 335 have been analyzed by Bentz et al. (2009a; 2013) and the one for
IRAS F12397+3333 has been treated by Mathur et al. (2012) who followed the
same scheme described in Bentz et al. (2009a). Our results are consistent with
both these earlier studies.

Figures $2-4$ show mean observed spectra, light curves and CCFs for the three 
sources.  
As described below, we found that
moon phases and the angular separation between the Moon and the target are the
main factors affecting the $V-$band light curve. We demonstrate this in Fig.~2
for one of the sources (Mrk 335) and do not use these light curves any more
since they are inferior to the $F_{5100}$ light curves in all three sources.
The variability characteristics, $F_{\rm var}$, calculated from the light
curves are given in Table 1. The averaged variability is typically $\sim 5\%$
over the monitoring campaign. The largest peaks or dips are significantly
larger than this value. For Mrk 335, Mrk 142 and IRAS F12397+3333, they show
$F_{5100}$ and H$\beta$ variations of
$(F_{5100}, {\rm H}\beta)\approx$ $(15\%,10\%)$, $(25\%,20\%)$, $(15\%,10\%)$,
respectively. All values are significantly larger than the uncertainties
associated with our line and continuum measurements.

The lags computed from the CCFs of the H$\beta$ emission line versus the $F_{5100}$
light curves are tabulated in Table 4 and discussed in \S3.2. We prefer the use of
$F_{5100}$ over the $V$-band since the latter may be influenced by the strong
emission lines of H$\beta$ and \feii\, that can slightly affect the lag
determination. The monochromatic luminosities, $L_{5100}$, calculated after allowing for 
Galactic extinction based on Schlafly \& Finkbeiner (2011), are listed in Table
5. We used the centroid time lag ($\tau_{\rm cent}$) to indicate our best value
of $\rblr$ that enters the calculation of the BH mass.

Our BH mass measurements are based on the FWHM of the broad H$\beta$ lines as measured from 
the mean spectra. The subtraction of the narrow H$\beta$ component is rather uncertain because 
of the very smooth line profiles and the limited spectral resolution of
our observations (about 500 km/sec based on a comparison with the SDSS spectra).  
Similar to the approach used in Hu et al. (2012), we fitted the entire H$\beta$ profile 
with a narrow Gaussian component ($F_{\rm N}$) and a broad Gaussian-Hermite component ($F_{\rm B}$). 
For a typical AGN with luminosity similar to the ones measured in our sample,  
$F_{\rm N}/F_{\rm [O~III]} \simeq 0.1$, where $F_{\rm [O~III]}$
is the flux of the narrow \oiii$\lambda5007$ line. We used this ratio, and the shift and FWHM of 
the \oiii$\lambda5007$ line, to define
$F_{\rm N}$. This was subtracted from the total profile to obtain the measured FWHM(broad H$\beta$). 
The uncertainty is obtained by following the same procedure assuming a) $F_{\rm N}=0$ and b) 
$F_{\rm N}=0.2F_{\rm [O~III]}$. Finally we obtained the intrinsic FWHM(broad H$\beta$) by allowing 
for the instrumental broadening. The FWHMs and their uncertainties are listed in Table 5.

Regarding the preferred velocity measure, there are two options: $\vfwhm$ and
$\sigma_{\rm line}$ (see \S1 for definitions and references). The mean values for both
are obtained through a comparison with the $\mbh-\sigma_*$ relationship in non-active
galaxies. The most recent results based on 25 AGNs with reliably measured $\sigma_*$
are those of Woo et al (2013) who obtained $\fblr(\sigma_{\rm line})=5.1-5.9$, depending 
on the exact method used. The uncertainty on these numbers are about $\pm 1.5$. Netzer and 
Marziani (2010) used the $\vfwhm$-based method and the somewhat inferior samples of Onken 
et al. (2004) and of Woo et al. (2010) to obtain $\fblr(\vfwhm)=1.0$.
This is confirmed by the more recent results of Woo (2013, private communication) 
that provides $\fblr(\vfwhm)$ for the above sample of 25 AGNs. The results of this work is
$\fblr(\vfwhm)=0.98^{+0.28}_{-0.22}$, i.e. similar to the scatter of $\fblr(\sigma_{\rm line}$).
Since for a Gaussian line profile $\vfwhm \simeq 2.35 \times \sigma_{\rm line}$, the two methods 
are basically identical given the uncertainties.

The RMS spectra over the region of the H$\beta$ line are shown in Figure 
5 where they are compared with the same parts of the mean spectra. The quality of the RMS spectra 
depends on the  seeing and the position of the target in the slit. The spectra tend to be noisy in 
particular in cases of small variability amplitude. Strong residuals may be present depending on 
the exact positioning of the slit and the accuracy of the wavelength calibration. In the cases 
studied here, this is most noticeable for the \oiii$\lambda$5007 line, the strongest narrow feature 
in the spectrum, where the residual noise, 
at a level of about 3\% of the line intensity, is clearly visible. This phenomenon is well known 
from earlier reverberation mapping campaigns (e.g. Figures 1-3 in Peterson et al. 1998, 
Kaspi et al. 2000, Park et al. 2012). 
We have also checked the measured fluxes of the \oiii$\lambda$5007 lines in all spectra and found 
that they fluctuated in a random way, uncorrelated with the continuum variations,  by 3\%, 4\% and 
4\% for Mrk 335, Mrk 142 and IRAS F12397, respectively. 

We measured  FWHM(H$\beta$) from the RMS spectra after smoothing the profiles with a nine 
pixels boxcar filter. The uncertainty on the width was estimated by repeating the procedure with a 
3-pixel boxcar filter that resulted in a narrower profile. These measurements resulted in FWHMs of
$1418\pm118$, $1623\pm110$ and $1510\pm194$ \kms, for Mrk 335, Mrk 142 and IRAS F12397+3333, 
respectively. The uncertainties derived in this way are only due to the noisy RMS profiles.
The comparison with the FWHMs measured from the mean spectra suggests a somewhat narrower RMS 
profile for Mrk 335, and similar witdthes, within the uncertainties, for the other two objects.

In this work we adopt the combination of the $\vfwhm$ method and the mean spectrum for several 
reasons: First, the measurement of $\vfwhm$, given our spectral resolution, is less uncertain than 
$\sigma_{\rm line}$. Second, NLS1s tend to have Lorentzian-shaped line profiles that can result 
in a considerable increase of $\sigma_{\rm line}$ which is very sensitive to the accuracy of the 
very extended wings measurement. Finally, in many cases the combination of the RMS spectrum and 
$\sigma_{\rm line}$ gives similar BH mass to the combination of $\vfwhm$ and the mean spectrum 
(Kaspi et al. 2000). This is not surprising given that the scaling factor $\fblr$ is an average 
over a large number of BLRs, with different radial gas distributions and inclinations to the lines 
of sight, some described better by $\vfwhm$ and some by $\sigma_{\rm line}$. 
%There is also a more fundamental issue regarding the accuracy of the line width measurement from 
%the RMS spectrum given that the line width is measured from the variable part of the line while 
%the time lag is measured by using the entire line intensity.
Regardless of the exact method, 
the only secure way to obtain the correct mass and its normalization is by calibrating the entire 
sample against the $\mbh-\sigma_*$ relationship.

The results of the three BH mass measurements are listed in Table 5 with their associated
uncertainties. We also list lower limits on Eddington ratios that were computed by adopting
the most conservative estimate on the radiation efficiency (see discussion below).
All three sources are super-Eddington accretors.

\subsection{Notes on individual objects}

\subsubsection{Mrk 335}
Mrk 335 is a well-known NLS1. It has been monitored for 6 years by
Kassebaum et al. (1997) and re-analyzed by Zu et al. (2011).
The X-ray observations show large variations at soft and hard X-ray energies
(Ar\'evalo et al. 2008). The optical monitoring  shows variability at a level of
10\% (Peterson et al. 1998). Figure \ref{mrk335} shows the line and continuum light
curves obtained by us with variability amplitudes of $\sim 15\%$ in both. The measured
lag, $10.6^{+1.7}_{-2.9}$ days, is similar to the previous measurements that range
from 12.5 to 16.8 days (see  Table 6 for details and references). This lag is still in
reasonable agreement with the Bentz et al. (2013) expression for the $\rblr-L$
relationship. The virial mass of the black hole measured by us is $8.3\times 10^6\sunm$
and the Eddington ratios $\mdotss\approx 1.6$, large enough to indicate a SEAMBH. The more
conservative approach assuming $\etaSS=0.038$, gives $\mdotss\approx 0.6$ which
is also beyond the SS regime of thin ADs (Laor \& Netzer 1989).

Comparison with previous RM experiments for
this source, listed in Table 6, suggest that the present response of the BLR is
shorter than in all previous campaigns although the deviation is still consistent
with the estimated uncertainties.
We suggest that the size of the BLR in this source changed significantly,
over a period of about 10 years since the earlier RM measurements. The
corresponding change in the 5100\AA\ luminosity is smaller.
This interesting possibility requires more data to instigate in detail.

\subsubsection{Mrk 142}
This object was mapped by the LAMP project (Bentz et al. 2009b) who reported a time lag of
$\tau=2.7\pm 0.8$ days measured from the peak correlation coefficient
($r_{\rm max}\approx 0.5$).
The significance of this measurement was the lowest among all sources monitored in that project.
The reason was the lack of a clear peak in the \hb\ light curve.

Figure \ref{mrk142} shows the light curves of Mrk 142 obtained by the SLT. Our observations are
superior to the measurements of Bentz et al. (2009b) for several reasons: the sampling is more
homogeneous, the error bars are smaller, and the light curves exhibit two clear minima
and maxima. The CCF shows a very sharp peak around $\tau_{\rm cent}\approx 6.4$  days with
a correlation coefficient of $r_{\rm max}\approx 0.7$. Our measured lag is larger than the
previously measured value by more than a factor 2 which brings the source closer to the
$\rblr-L$ relationship of Bentz et al. (2013). However, the object is still a clear outlier
of the relationship.

According to Eqn.~3 and Table 4, the value of $\mdotss$ derived for Mrk 142 is at least 2.3
($\etaSS=0.038$) and more likely around 5.9 ($\etaSS=0.1$).
This is at the high end of the $\mdotss$ distribution of all AGNs with reliable RM-based BH
mass measurements. As alluded to earlier, such accretion rates are inconsistent with the
thin disk idea and we expect that the real accretion
rate, $\mdot$, is even larger making Mrk 142 a clear case of a SEAMBH.

\subsubsection{IRAS F12397+3333}
IRAS F12397+3333 is an infrared luminous source
identified as an AGN by Keel et al. (1988). The optical SED of the source
is  ``red'' compared with most AGNs, including those in the RM sample. This is
typical of other IRAS sources. The hypothesis of significant dust attenuation is supported by
polarization measurements (Grupe et al. 1998) and by the weak signal obtained
by {\it GALEX} at $\lambda=1528$\AA\ (Atlee \& Mathur 2009) which, when combined
with our own measurements, gives $L_{5100}/L_{1528} \simeq 4-8$, far above what
is observed in unreddened AGNs, like those in the SDSS sample (note that these
are not contemporaneous observations).

To further check this point, we fitted the SDSS spectrum of the source and decomposed
the H$\alpha$ and H$\beta$ lines into narrow and broad components\footnote{The spectral
resolution of the YFOSC for a $2.^{\prime\prime}5$ slit is $\lambda/\Delta\lambda \sim 600$,
which is too low to separate the narrow and broad components. This is the reason for
using the higher resolution SDSS spectrum.}. We found FWHMs of 380 and 1680 \kms, for
the narrow and broad components, respectively. This results in
$\left({\rm H\alpha/H\beta}\right)_{\rm broad}\approx 5.71$ and
$\left({\rm H\alpha/H\beta}\right)_{\rm narrow}\approx 6.14$.  This steep Balmer
decrement suggests that
reddening significantly affects the narrow emission lines and possibly also the broad
emission lines.  Finally, the X-ray spectrum of the source shows strong and
broad K$\alpha$ line emission at around 6.4 keV (Bianchi et al. 2009; Zhou \& Zhang
2010) but the hydrogen column density is low, $N_{\rm H} = (2.74\pm
1.38)\times 10^{20}{\rm cm^{-2}}$ (Grupe et al. 2001).  Assuming galactic
dust-to-gas ratio, this corresponds to $A_v \simeq 0.15$.

Given the likely continuum extinction, and the conflicting results obtained from the
X-ray column density and the {\it GALEX} results, we considered two different
possibilities regarding the reddening. The first is no attenuation but extreme value
of ${\rm H\alpha/H\beta}$ in the BLR and the second is dust extinction with an intrinsic
Balmer decrement of $\left({\rm H\alpha/H\beta}\right)_{\rm broad} = 3.5$, a ratio typical
of a large number of ``blue'' AGNs with no suspected continuum attenuation (e.g., Vanden
Berk et al. 2001). Balmer decrements in this range 
are fairly typical and are
consistent with the conditions expected in the BLR (Netzer 2013).

Assuming first no reddening, we get $L_{5100}=3.9\times 10^{43}\ergs$. Using Eqn. 3
and our measured BH mass ($7.5\times 10^6\sunm$) we get $\mdotss \simeq 1.3$. In the
second case, $L_{5100}=1.69 \times 10^{44}\ergs$ which results in $\mdotss \simeq 12.1$.
The corresponding values for $\etaSS=0.038$ are 0.5 and 4.6. All values are above or close
the limiting case of $\mdotss=1$ and the ones corrected for reddening are the largest
among all cases published so far in AGNs with directly measured BH mass.
Comparing Eddington ratio estimations in other samples of NLS1s with estimates
based on the virial method, e.g. the one shown in Figure 4 of  Wang \& Netzer 2003,
we find that IRAS F12397+3333 and Mrk 142 have the highest Eddington ratios measured
so far.

\subsection{BH mass, Eddington rates, growth times and cosmological implications}

As argued in numerous publications (e.g., Kelly et al. 2011 and references therein),
understanding AGN variability is key to the understanding of the accretion mechanism,
including the AD properties (Liu et al. 2008). Thus, the results of our monitoring
campaign shed new
light on several aspects of NLS1s in general and SEAMBHs in particular. First, we
detected significant line and continuum variations, $\sim 15\%$, in three fast
accreting black holes. Such variations have not been detected in most other known
cases of NLS1s. The variations can provide important constraints on the properties
of slim accretion disks (Mineshige et al 2000), a topic that we are currently
investigating.

Second, all three objects are accreting close to or above the limit of
$\mdotss\approx 1$ even under the most conservative estimate of $\etaSS=0.038$.
Two of the sources, Mrk 142 and IRAS F12397+3333, show the highest values
of $\mdotss$ measured so far. They indicate physical conditions that are far
outside the nominal conditions for thin SS type disks. Among reverberation-mapped
AGNs, these are perhaps the best candidates for having slim accretion disks.

Third, the newly measured lag for Mrk\,335 is smaller than the earlier measurements but 
still in reasonable agreement with the $\rblr-L$ relation obtained by Bentz et al. (2013).
This is not the case for Mrk\,142, and for IRAS F12397+3333 assuming there is extinction
along the line of sight to this source. Both these objects fall well below the
above correlation.
This leads to a suggestion that SEAMBHs may exhibit a somewhat different
$\rblr-L$ relationship. One possibility is that in such cases, the very different
geometry of the slim disk results in a different radiation pattern.
In particular, the face-on luminosity can be very large
compared with the equatorial value leading to the possibility that equatorial BLR
gas clouds are exposed to much weaker radiation than assumed in isotropic radiation 
fields, which affect their location. Such
ideas have been around for some time (Netzer 1987; Korista, Ferland \& Baldwin 1997;
Netzer 2013) but have never been studied in relation to slim disks. In addition,
the balance between gravity and radiation pressure force, which is proportional to
$\mdot$, may affect BLR gas around slim disks in SEAMBHs more than in other AGNs.
This can lead to marginally bound, or perhaps even
unbound cloud systems (see discussion of these ideas in Marconi et al. 2008 and
Netzer and Marziani 2010). If correct, the BH mass determination in such sources
may be less secure. On the other hand, 7 of the sources with measured  $\sigma_*$
studied by Woo et al (2013) are NLS1s
and their virial-based BH mass is in reasonable agreement with the mass obtained
by the $\mbh-\sigma_*$ relationship. Unfortunately none of the three sources studied
here is included in the Woo et al. (2013) sample.

Fourth, the normalized accretion rates obtained here, at least for two of the
cases, are so large that they must
be important to the general issue of BH growth and BH duty cycle. In particular,
in the exponential growth scenario,  the growth time of massive BHs is inversely
proportional to $\mdot$. Typical continuous growth times for local active BHs,
based on typical Eddington rates of such sources  ($ \sim 0.1$), are of order
the Salpeter time, $\sim 4 \times 10^8 $ yr (e.g., Netzer \& Trakhtenbrot
2007). For the objects considered here, the growth time could be an order of
magnitude or much shorter (see Netzer and Trakhtenbrot 2013 for similar
considerations regarding growth times of SDSS AGNs).

Finally, the direct measurements of SEAMBH masses will allow us to estimate
the cosmological distance to such objects using the relationship between the
BH mass and the saturated luminosity of such objects (Wang et al. 2013). This
can provide a new way to measure cosmological distances up to very high
redshifts. We are extending the monitoring project to larger numbers and to
higher redshifts in order to provide more precise calibration of this
 method.

\section{Conclusions}

We introduced the first results from a new  project aimed at measuring accurate
masses in AGNs hosting the fastest accreting active BHs. The first stage of the
project includes 10 targets that were monitored
with the Shangri-La telescope at the Lijiang station of the Yunnan observatory.
The results  pertain to three NLS1s for which we obtained detailed light curves
and meaningful CCFs that can be used to derive accurate time lags and reliable
BH masses. We find  H$\beta$ time lags of $10.6_{-2.9}^{+1.7}$,
$6.4_{-2.2}^{+0.8}$ and $11.4_{-1.9}^{+2.9}$ days for Mrk 335, Mrk 142 and
IRAS F12397, respectively. For Mrk 142 and IRAS F12397 (assuming its continuum is
extincted), the lags are shorter than expected from the most recent $\rblr-L$
relationships for the general AGN population. The corresponding BH masses are
$(8.3_{-3.2}^{+2.6})\times 10^6\sunm$, $(3.4_{-1.2}^{+0.5})\times 10^6\sunm$
and $(7.5_{-4.1}^{+4.3})\times 10^6\sunm$ and the corresponding Eddington ratios
($\mdotss$) 1.6, 5.9 and 12.1 (extincted) or 1.3 (unextincted). All these values
assume $\etaSS= 0.1$. Values of $\dot{m}$ as high as those measured for Mrk 142 and
IRAS F12397 have never been directly measured in other AGNs. All  three  BHs are undergoing
super-Eddington accretion with important consequences to the BH accretion mechanism,
BH growth rate and, perhaps, cosmology.

\acknowledgements{%
We acknowledge stimulating discussions with L. C. Ho and useful comments by the referee.
We are very grateful to the staff
of the Lijiang Station of the Yunnan Observatory for their enthusiasm and endless support
and for operating the Shangri-La telescope so efficiently and professionally. Without
their great effort, the project could not have been completed. We thank K. Fang for helping
in the observations during May 2013, and N. Jiang for useful suggestions iabout the HST image 
analysis. This research is supported by the following grants:
the Special Funds for Major State Basic Research Projects (the 973 project) through 2009CB824800,
NSFC-11173023, -11233003 and -11133006, and the China-Israel ISF-NSFC grant 83/13.}

\newpage

\begin{deluxetable}{llllllccccccc}
\rotate
\tablecolumns{7}
\tablewidth{0pc}
\tablecaption{Basic data and variability amplitude}
\tabletypesize{\footnotesize}
\tablehead{
\colhead{Object}                &
\colhead{$\alpha_{2000}$}       &
\colhead{$\delta_{2000}$}       &
\colhead{redshift}              &
\colhead{$E$(B-V)}              &
\colhead{monitoring period}     &
\colhead{$N_{\rm spec}$}        &
\multicolumn{3}{c}{variability amplitude($\%$)$^*$}& &
\multicolumn{2}{c}{Comparison stars}\\ \cline{8-10}\cline{12-13}
\colhead{}                      &
\colhead{}                      &
\colhead{}                      &
\colhead{}                      &
\colhead{}                      &
\colhead{}                      &
\colhead{}                      &
\colhead{$F_{\rm var}$(5100\AA)}&
\colhead{$F_{\rm var}$(V)}      &
\colhead{$F_{\rm var}({\rm H\beta}$)}& &
\colhead{$R_*$}           &
\colhead{P.A.}
}
\startdata
Mrk 335	   &00 06 19.5&+20 12 10&0.0258&0.030 &2012 Oct$-$2013 Feb& 91 &5.2$\pm$0.5 &3.9$\pm$0.3 &3.4$\pm$0.3 && $ 80^{\pp}.7$&174.5$^{\circ}$\\
Mrk 142	   &10 25 31.3&+51 40 35&0.0449&0.015 &2012 Nov$-$2013 Apr&119 &8.1$\pm$0.6 &5.5$\pm$0.4 &7.8$\pm$0.5 && $113^{\pp}.1$&155.2$^{\circ}$\\
IRAS F12397+3333&12 42 10.6&+33 17 03&0.0435&0.017 &2013 Jan$-$2013 May& 51 &5.6$\pm$0.6 &3.2$\pm$0.4 &4.5$\pm$0.6 && $189^{\pp}.0$&130.0$^{\circ}$\\
\enddata
\tablecomments{$E(B-V)$ is the Galactic extinction using the maps in Schlafly \& Finkberiner (2011). 
$N_{\rm spec}$ is the number of spectroscopic observing epochs, $R_*$ is the angular
distance to the target and P.A. is the position angle. $^*$Amplitudes were calculated using Eqn.~4. The
uncertainties of $F_{\rm var}$ is calculated according to Edelson et al. (2002).
}
\end{deluxetable}

%\clearpage

\begin{deluxetable}{cccccccccccc}
\rotate
\tablecolumns{9}
\tablewidth{0pc}
\tablecaption{Continuum and H$\beta$ light curves}
\tabletypesize{\footnotesize}
\tablehead{
\multicolumn{3}{c}{Mrk 335} & & \multicolumn{3}{c}{Mrk 142} & & \multicolumn{3}{c}{IRAS F12397+3333} \\
\cline{1-3} \cline{5-7} \cline{9-11}
\colhead{JD} & \colhead{$F_{5100}$} & \colhead{$F_{\rm H\beta}$} & &
\colhead{JD} & \colhead{$F_{5100}$} & \colhead{$F_{\rm H\beta}$} & &
\colhead{JD} & \colhead{$F_{5100}$} & \colhead{$F_{\rm H\beta}$} }
\startdata
22.1913 & $ 6.120\pm 0.118$ & $ 5.904\pm 0.017$ & &37.4339 & $ 1.552\pm 0.010$ & $ 0.800\pm 0.007$ & &115.4571 & $ 1.982\pm 0.009$ & $ 0.789\pm 0.007$ \\
23.3037 & $ 6.478\pm 0.040$ & $ 6.231\pm 0.033$ & &54.4214 & $ 1.701\pm 0.010$ & $ 0.798\pm 0.006$ & &116.4462 & $ 2.069\pm 0.008$ & $ 0.804\pm 0.006$ \\
24.0274 & $ 6.285\pm 0.023$ & $ 6.002\pm 0.021$ & &55.4262 & $ 1.811\pm 0.019$ & $ 0.807\pm 0.008$ & &150.4556 & $ 1.973\pm 0.017$ & $ 0.820\pm 0.011$ \\
25.0172 & $ 6.056\pm 0.016$ & $ 5.814\pm 0.017$ & &59.4444 & $ 1.619\pm 0.012$ & $ 0.825\pm 0.006$ & &151.4504 & $ 1.936\pm 0.014$ & $ 0.769\pm 0.010$ \\
26.0242 & $ 6.167\pm 0.021$ & $ 5.797\pm 0.026$ & &60.4155 & $ 1.631\pm 0.009$ & $ 0.823\pm 0.008$ & &152.4339 & $ 1.966\pm 0.022$ & $ 0.813\pm 0.009$ \\
\enddata
\tablecomments{The full version of this table is also available in machine-readable form in the electronic edition of the
{\it Astrophysical Journal}. JD: Julian dates from 2456200; $F_{5100}$ and $F_{\rm H\beta}$ are fluxes at $(1+z)5100$\AA\, and H$\beta$
emission line in units of $10^{-15}\mathrm{~erg~s^{-1}~cm^{-2}~\AA^{-1}}$ and $10^{-13}\mathrm{~erg~s^{-1}~cm^{-2}}$, respectively.
The systematic uncertainties of $F_{5100}$ and $F_{\rm H\beta}$ are
($\Delta F_{5100}$, $\Delta F_{\rm H\beta})= (0.138, 0.091), (0.045, 0.018)$ and $(0.035, 0.018)$
 for Mrk 335, Mrk 142 and IRAS F12397 respectively.}
\end{deluxetable}

\begin{deluxetable}{lllccccccc}
\tablecolumns{10}
\tablewidth{0pc}
\tablecaption{Host galaxy decomposition}
\tabletypesize{\footnotesize}
\tablehead
{
\colhead{Object} & \colhead{Data set} & \colhead{Observational setup} & \colhead{$m_{\rm st}^*$} &
\colhead{$R_e$} & \colhead{$n$} & \colhead{$b/a$} & \colhead{P.A.} &
\colhead{Note} & \colhead{$\chi_{\nu}^2$} \\
 & & & & ($''$) & & & (deg) & &
}
\startdata
Mrk 335          & J9MU010 & ACS, HRC, F550M  & 14.73  &      &       &      &       & PSF             & 1.700 \\
                 &         &                  & 17.78  & 0.09 & 0.89  & 0.04 & 44.45 & Add'l PSF       &       \\
                 &         &                  & 14.81  & 5.02 & 3.58  & 0.93 & 95.81 & Elliptical      &       \\
                 &         &                  & -0.007 &      &       &      &       & Sky             &       \\
Mrk 142          & IB5F010 &WFC3, UVIS1, F547M& 16.13  &      &       &      &       & PSF             & 1.424 \\
                 &         &                  & 18.27  & 0.41 & 0.45  & 0.78 & 14.25 & Bulge           &       \\
                 &         &                  & 16.34  & 4.73 & [1.0] & 0.56 & 39.53 & Disk            &       \\
                 &         &                  & 0.014  &      &       &      &       & Sky             &       \\
IRAS F12397      & J96I090 & ACS, HRC, F625W  & 16.27  &      &       &      &       & PSF             & 1.630 \\
                 &         &                  & 18.36  & 0.16 & 1.16  & 0.72 & 42.65 & Nuclear Spiral? &       \\
                 &         &                  & 17.40  & 0.79 & 0.75  & 0.96 & 146.4 & Bulge           &       \\
                 &         &                  & 16.51  & 3.69 & [1.0] & 0.49 & 53.84 & Disk            &       \\
                 &         &                  & 0.007  &      &       &      &       & Sky             &       \\
\enddata
\tablecomments{The values in square brackets are fixed in the fitting procedure.
*$m_{\rm st}$ is the ST magnitude, an $f_{\lambda}$-based magnitude system,
$m_{\rm ST} = -2.5 \log_{10}(f_\lambda)-21.10$, for $f_\lambda$ in erg~s$^{-1}$cm$^{-2}$\AA$^{-1}$
(see Sirianni et al. 2005). The units of sky are electrons/s.}
\end{deluxetable}

\begin{deluxetable}{llll}
\tablecolumns{8}
\tablewidth{0pc}
\tablecaption{Results of CCF analysis in rest frame of sources}
\tabletypesize{\footnotesize}
\tablehead{                                        &
\multicolumn{3}{c}{\raisebox{1ex}{$F_{5100}{\rm~versus~H\beta}$}} \\ \cline{2-4}
\colhead{Object}            &
\colhead{$\tau_{\rm cent}$} &
\colhead{$\tau_{\rm peak}$} &
\colhead{$r_{\rm max}$}     \\
\colhead{}&
\colhead{(days)}&
\colhead{(days)}
}
\startdata
Mrk 335	         & $10.6_{-2.9}^{+1.7}$ & $8.2_{-1.1}^{+4.4}$  & 0.67  \\
Mrk 142	         & $6.4_{-2.2}^{+0.8}$  & $5.3_{-2.2}^{+2.6}$  & 0.68   \\
IRAS F12397+3333 & $11.4_{-1.9}^{+2.9}$ & $11.9_{-2.2}^{+1.4}$ & 0.71  \\
\enddata
\end{deluxetable}

\begin{deluxetable}{lccccccc}
\tablecolumns{8}
\tablewidth{0pc}
\tablecaption{Luminosity, BH mass and Eddington ratios}
\tabletypesize{\footnotesize}
\tablehead{
\colhead{Object}            &
\colhead{FWHM$^a$}          &
%\colhead{$F_{\rm obs}^b$}   &
\colhead{$F_{\rm gal}^b$}     &
\colhead{$F_{\rm AGN}^b$}     &
\colhead{$L_{5100}$} &
\colhead{$\mbh^c$ }       &
\multicolumn{2}{c}{$\mdotss$} \\ \cline{3-4}\cline{7-8}
\colhead{}                  &
\colhead{(\kms)}            &
\multicolumn{2}{c}{($10^{-15}~{\rm erg~s^{-1}~cm^{-2}~\AA^{-1}}$)}     &
%\colhead{($10^{-15}~{\rm erg~s^{-1}~cm^{-2}~\AA^{-1}}$)}     &
\colhead{($10^{43}\ergs)$}                  &
\colhead{($10^6\sunm$)}  &
\colhead{$\etaSS=0.1$} &
\colhead{$0.038$}
} %
\startdata
Mrk 335        & 1997$\pm$265 & 0.99$\pm$0.10 & 5.20$\pm$0.37 & 4.9$\pm$0.4 & $8.3_{-3.2}^{+2.6}$ & 1.6 & 0.6\\
Mrk 142        & 1647$\pm$ 69 & 0.37$\pm$0.04 & 1.27$\pm$0.15 & 3.7$\pm$0.4 & $3.4_{-1.2}^{+0.5}$ & 5.9 & 2.3\\
IRAS F12397(I) & 1835$\pm$473 & 0.65$\pm$0.06 & 1.44$\pm$0.14 &16.9$\pm$1.7 & $7.5_{-4.1}^{+4.3}$ &12.1 & 4.6\\
IRAS F12397(II)& 1835$\pm$473 & 0.65$\pm$0.06 & 1.44$\pm$0.14 & 3.9$\pm$0.4 & $7.5_{-4.1}^{+4.3}$ & 1.3 & 0.5\\
\enddata
\tablecomments{IRAS F12397(I)/(II) mean with/without intrinsic reddening, respectively.\\
$^a$FWHM stands for  $V_{\rm FWHM}$ in the text of the paper and is measured from the mean spectra.\\
 $^b$$F_{\rm gal}$ and $F_{\rm
AGN}$ are host galaxy and mean AGN fluxes at $(1+z)5100{\rm \AA}$. 
$F_{\rm gal}+F_{\rm AGN}=F_{\rm obs}$, where $F_{\rm obs}$ is the observed flux.\\
 $^c$$\mbh$ are
calculated using the centroid time lag, $V_{\rm FWHM}$ and $f_{\rm BLR}=1.0$. Uncertainties on 
 $\mbh$ are the results of the errors on the time lags and the FWHMs. 
$L_{5100}$ is the mean AGN luminosity at rest-frame wavelength of 5100\AA\ 
after the subtraction of the host galaxy contribution and allowing for Galactic extinction.
}
\end{deluxetable}

\begin{deluxetable}{cccccl}
\tablecolumns{6}
\tablewidth{0pc}
\tablecaption{Mrk 335: BLR size and continuum luminosity at different epochs}
\tabletypesize{\footnotesize}
\tablehead{
\colhead{Epoch} &
\colhead{$\tau_{\rm cent}$} &
\colhead{$\tau_{\rm peak}$} &
\colhead{$F_{\rm AGN}[5100{\rm \AA}(1+z)]$ } &
\colhead{$L_{5100}$ } &
\colhead{Reference}\\
\colhead{(JD2400000+)}  &
\colhead{(days)}        &
\colhead{(days)}        &
\colhead{(${\rm 10^{-15}~erg~s^{-1}~cm^{-2}~\AA^{-1}}$)}&
\colhead{(${\rm 10^{43}erg~s^{-1}}$)} }
\startdata
49156$-$49338   & $16.8^{+4.8}_{-4.2}$ & $18^{+5}_{-6}$       & $6.12\pm0.19$  & $5.8\pm0.2$ & P04, B13   \\
49889$-$50118   & $12.5^{+6.6}_{-5.5}$ & $13^{+9}_{-7}$       & $7.25\pm0.21$  & $6.8\pm0.2$ & P04, B13   \\
55440$-$55568   & $14.3^{+0.7}_{-0.7}$ & $14.0^{+0.9}_{-0.9}$ & $5.84\pm0.29$  & $5.5\pm0.3$ & G12, B13   \\
56222$-$56328   & $10.6^{+1.7}_{-2.9}$ & $8.2^{+4.4}_{-1.1}$  & $5.20\pm0.37$  & $4.9\pm0.4$ & this paper\\
\enddata
\tablecomments{P04: Peterson et al. (2004); G12: Grier et al. (2012); B13: Bentz et al. (2013)
}
\end{deluxetable}

\clearpage

\begin{figure*}[t!]
\begin{center}
\includegraphics[angle=0,width=0.95\textwidth]{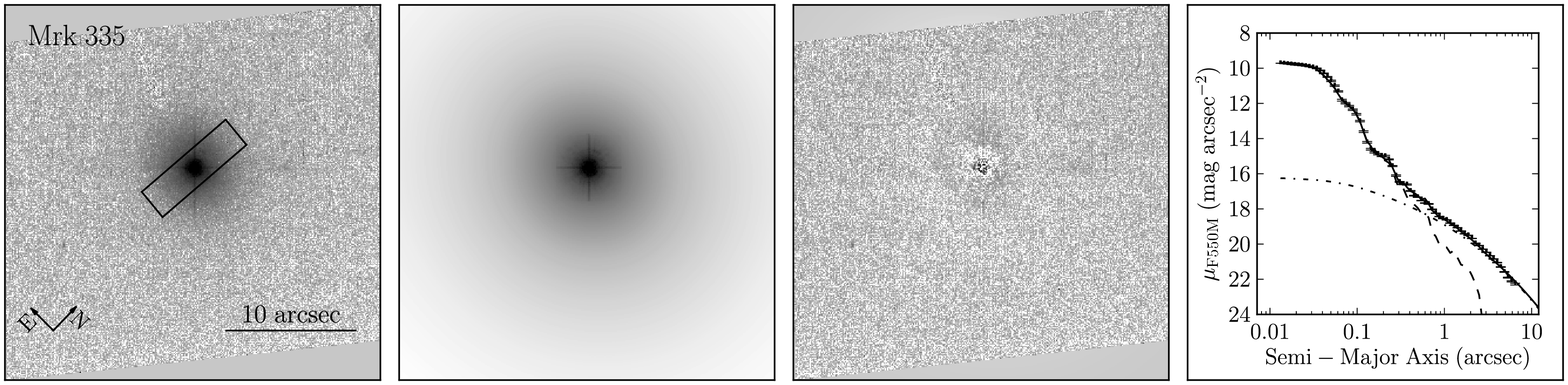}
\includegraphics[angle=0,width=0.95\textwidth]{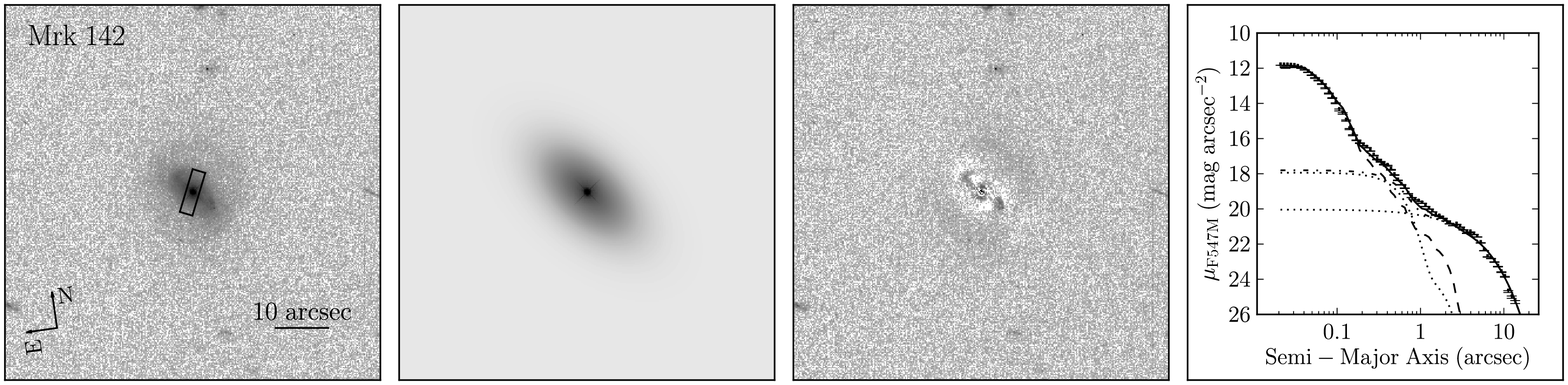}
\includegraphics[angle=0,width=0.95\textwidth]{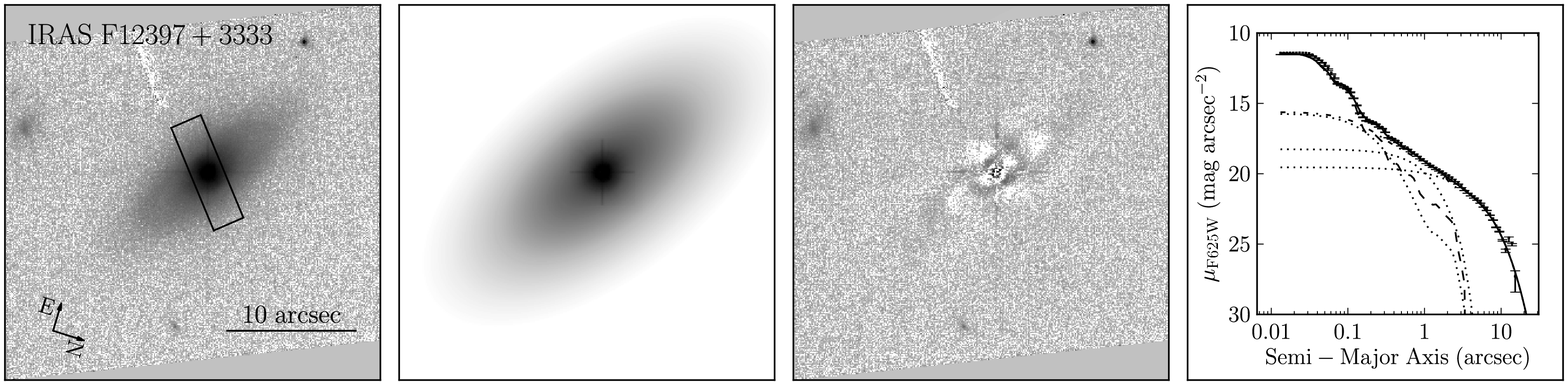}
\end{center}
\caption{\footnotesize
Hubble Space Telescope images of Mrk 335, Mrk 142 and IRAS F12397+3333. The
left panels show the original images and the small boxes illustrate the
spectroscopic aperture used to extract the spectrum.
The 2nd column shows model images, the 3rd one  the residuals obtained
after subtracting the fitted model. The
4th column shows one-dimensional surface profiles of the three galaxies.
Points with error bars are observed data, solid lines are the best-fit models,
dashed lines are PSFs, dash-dotted lines are host
profiles and dotted lines are the components (S\'ersic profiles) used to model the host galaxy light.
}
\label{mrk335_lc}
\end{figure*}

\clearpage

\begin{figure*}[t!]
\begin{center}
\includegraphics[angle=0,width=0.9\textwidth]{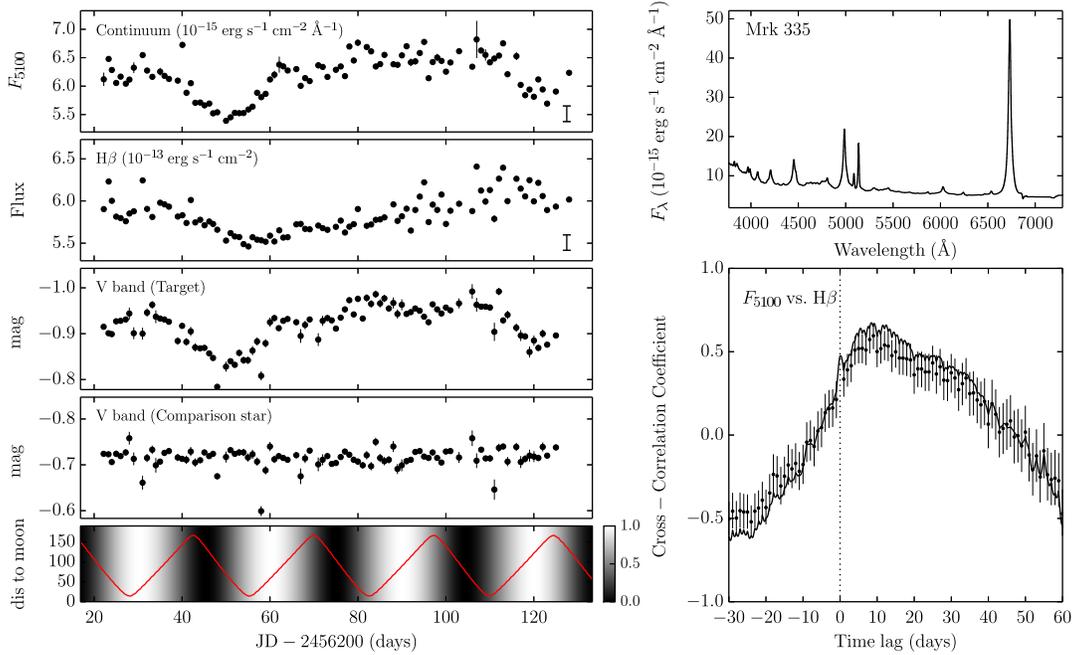}%{Mrk335_tot.eps}
\end{center}
\caption{\footnotesize
Results for Mrk 335. From top to bottom and from left to right:
1. $F_{5100}$ light curve. 2.  H$\beta$ light curve. 3. $V-$band light curve 4. $V-$band light curve
of the comparison star. 5. Moon phases and angular separation, in degrees, between the target and the
moon (red line). The Moon brightness is coded on the right side of the panel (1.0 is full moon). 6.
Mean observed spectrum. 7. The CCF of $F_{5100}$$-$H$\beta$ light curves.
Points with error bars are from the ZDCF method and solid line from the ICCF method.
Note the effect of the moon on the comparison star magnitude. The bars plotted in the
right lower corners of $F_{5100}$ and $F_{\rm H\beta}$ light curves are the
systematic uncertainties (see details in \S2.3.1).
}
\label{mrk335}
\end{figure*}

\begin{figure*}[t!]
\begin{center}
\includegraphics[angle=0,width=0.9\textwidth]{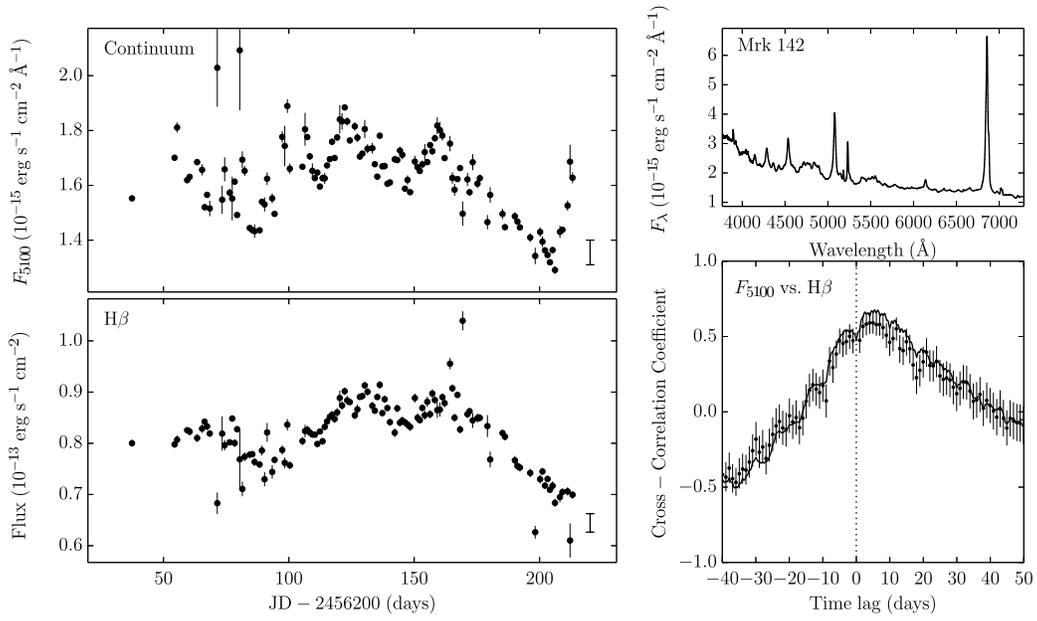}%{Mrk142_tot.eps}
\end{center}
\caption{\footnotesize
Results for Mrk 142. The left two panels are light curves of continuum at 5100\AA\, and H$\beta$, and
the right two panels are mean spectrum and CCF.
}
\label{mrk142}
\end{figure*}

\vspace{1cm}
\begin{figure*}[t!]
\begin{center}
\vspace{0.85cm}
\includegraphics[angle=0,width=0.9\textwidth]{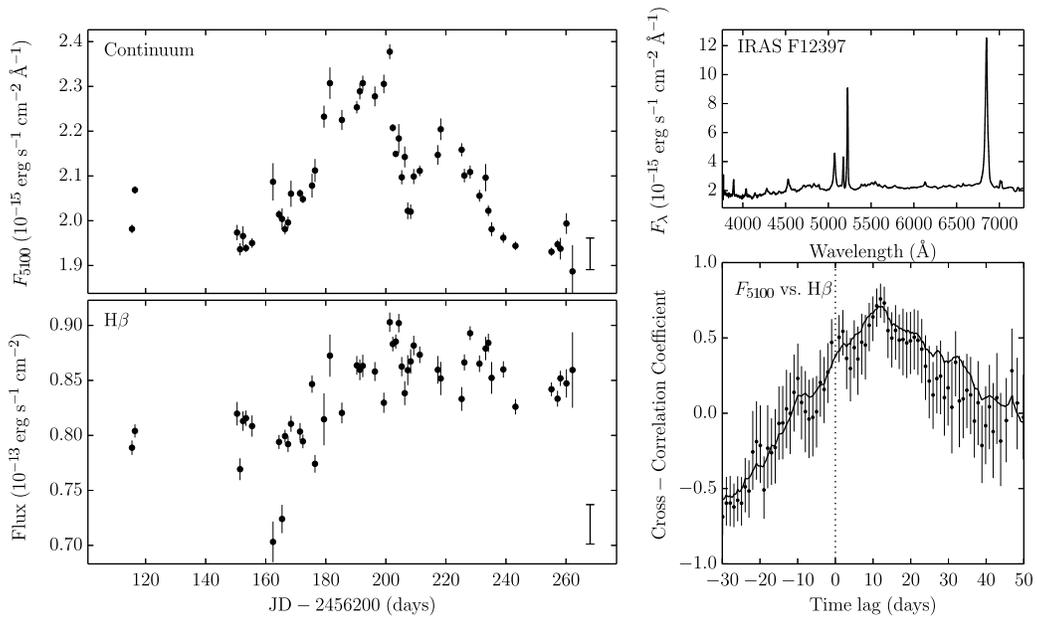}
\end{center}
\caption{\footnotesize
Results for IRAS F12397+3333 (same as Fig.~3).
}
\label{irasf12397}
\end{figure*}

\clearpage

\begin{figure*}[t!]
\begin{center}
\vspace{0.85cm}
\includegraphics[angle=0,width=\textwidth]{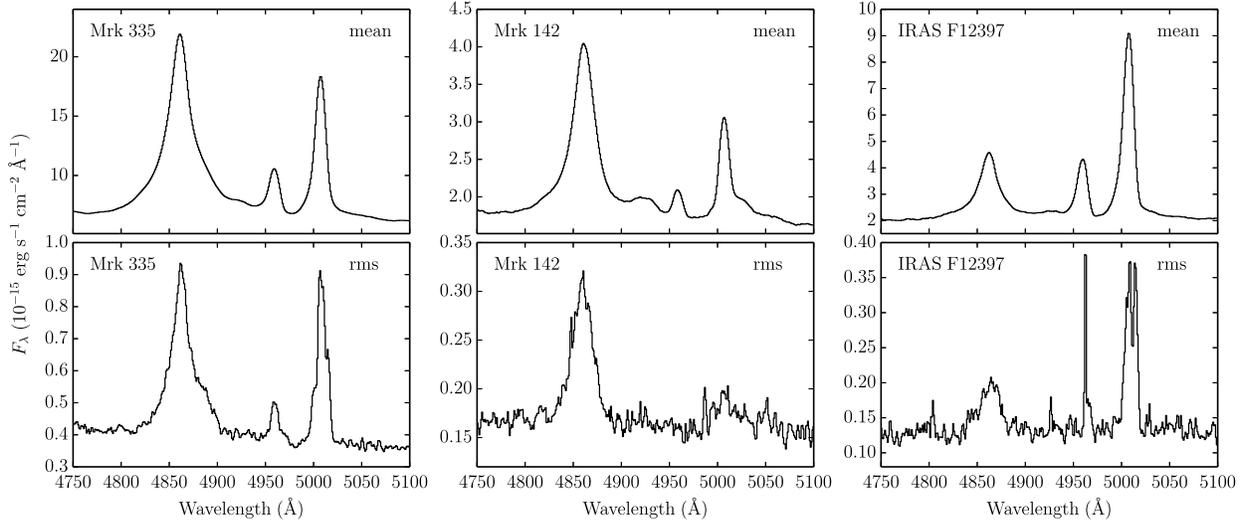}
\end{center}
\caption{\footnotesize
Mean and RMS spectra (observed flux vs. rest-frame wavelength)  of the three objects. The spectra 
are normalized to roughly the same vertical scale to enable a clearer comparison. Note the much 
noiser RMS spectrum and the similar line widths.}
\label{rms}
\end{figure*}

\end{document}